\documentclass[twocolumn]{aastex62}
\usepackage{graphicx,subfigure,epsfig,afterpage}
\usepackage{natbib}
\usepackage{appendix}
\usepackage{ulem}

\received{September 10, 2018}
\accepted{January 10, 2019}

\submitjournal{ApJ}

\shorttitle{Benchmarking Substellar Evolutionary Models}
\shortauthors{Wood et al.}

\begin{document}

\title{Benchmarking Substellar Evolutionary Models Using New Age Estimates for HD~4747~B and HD~19467~B}

\correspondingauthor{Charlotte M. Wood}
\email{cwood12@nd.edu}

\author[0000-0003-4773-4602]{Charlotte M. Wood}
\affiliation{University of Notre Dame, Department of Physics, 225 Nieuwland
Science Hall, Notre Dame, IN 46556}
\author[0000-0001-9879-9313]{Tabetha Boyajian}
\affiliation{Louisiana State University, Department of Physics and Astronomy, 202 Nicholson Hall, Baton Rouge, LA 70803}
\author[0000-0002-5823-4630]{Kaspar von Braun}
\affiliation{Lowell Observatory, 1400 W Mars Hill Rd, Flagstaff, AZ 86001}
\author[0000-0002-9873-1471]{John M. Brewer}
\affiliation{Yale University, Department of Astronomy, P.O. Box 208101,
New Haven, CT 06520}
\affiliation{Columbia University, Department of Astronomy, Mail Code 5246, 550 West 120th Street, New York, NY 10027}
\author[0000-0003-0800-0593]{Justin R. Crepp}
\affiliation{University of Notre Dame, Department of Physics, 225 Nieuwland
Science Hall, Notre Dame, IN 46556}
\author[0000-0001-5415-9189]{Gail Schaefer}
\affiliation{The CHARA Array of Georgia State University, Mount Wilson, CA 91023}
\author[0000-0002-7139-3695]{Arthur Adams}
\affiliation{Yale University, Department of Astronomy, P.O. Box 208101,
New Haven, CT 06520}
\author[0000-0002-6980-3392]{Timothy R. White}
\affiliation{Stellar Astrophysics Centre, Department of Physics and Astronomy, Aarhus University, Ny Munkegade 120, DK-8000 Aarhus C, Denmark}


\begin{abstract}
Constraining substellar evolutionary models (SSEMs) is particularly difficult due to a degeneracy between the mass, age, and luminosity of a brown dwarf. In cases where a brown dwarf is found as a directly imaged companion to a star, as in HD~4747 and HD~19467, the mass, age, and luminosity of the brown dwarf are determined independently, making them ideal objects to use to benchmark SSEMs. Using the Center for High Angular Resolution Astronomy Array, we measured the angular diameters and calculated the radii of the host stars HD~4747~A and HD~19467~A. After fitting their parameters to the Dartmouth Stellar Evolution Database, MESA Isochrones and Stellar Tracks, and Yonsei-Yale isochronal models, we adopt age estimates of $10.74^{+6.75}_{-6.87}$ Gyr for HD~4747~A and $10.06^{+1.16}_{-0.82}$ Gyr for HD~19467~A. Assuming the brown dwarf companions HD~4747~B and HD~19467~B have the same ages as their host stars, we show that many of the SSEMs under-predict bolometric luminosities by $\sim$ 0.75 dex for HD~4747~B and $\sim 0.5$ dex for HD~19467~B. The discrepancies in luminosity correspond to over-predictions of the masses by $\sim$ 12\% for HD~4747~B and $\sim$ 30\% for HD~19467~B. We also show that SSEMs that take into account the effect of clouds reduce the under-prediction of luminosity to $\sim 0.6$ dex and the over-prediction of mass to $\sim 8\%$ for HD~4747~B, an L/T transition object that is cool enough to begin forming clouds. One possible explanation for the remaining discrepancies is missing physics in the models, such as the inclusion of metallicity effects.
\end{abstract}



\section{Introduction}\label{sec:intro}
The atmospheres of brown dwarfs are quite complicated, including multiple convection zones, the possibility of cloud formation, and the presence of molecules that results in highly wavelength-dependent opacities \citep{mar15}. Atmospheric effects are also the main factor in determining how a brown dwarf evolves and cools. If we hope to fully understand brown dwarfs and other substellar objects, we need models that take into account all of these effects. Having complete models is especially important when studying free-floating ``field'' brown dwarfs, whose properties cannot be determined other than from the atmosphere.

Recent substellar evolutionary models do a better job at predicting optical color of brown dwarfs and matching observations for older objects than their predecessors \citep{bar15}. However, tests of these models are still fairly limited due to degeneracies between mass, age, and luminosity for brown dwarfs; a young, less massive brown dwarf can appear to have the same luminosity as an old, more massive brown dwarf. These degeneracies are the main source of uncertainty in age estimates for field brown dwarfs, inhibiting the accuracy of model tests. To properly constrain the models, we need benchmark brown dwarfs -- objects whose masses, ages, and luminosities can be determined independently.

The mass of a benchmark brown dwarf can be calculated using the orbital mechanics of the system in which it is found \citep{liu08, dup09b, cre11, dup17}. Other properties of a benchmark brown dwarf -- such as age and metallicity -- can be more readily inferred by studying the host star rather than the brown dwarf itself.

Using isochronal models, a more accurate age estimate of the host star can be determined by measuring the precise stellar radius, which places additional constraints on the location of the star on the HR-diagram \citep{cre12}. For nearby stars (d $\leq$ 50 parsecs), it is possible to determine the stellar radius precisely using interferometry \citep{boy12a, boy12b}.

In this paper, we present angular diameter measurements from the Center for High Angular Resolution Astronomy (CHARA) Array and calculate the radius (\S\ref{sec:observations}) of two Sun-like stars, HD~4747~A and HD~19467~A, known to host benchmark brown dwarf companions \citep{cre14, cre16, cre18}. We also present new age estimates for these systems (\S\ref{sec:ages}) using the Dartmouth Stellar Evolution Database, MESA Isochrones and Stellar Tracks (MIST), and Yonsei-Yale (YY) isochrone models \citep{dot07, dot08, pax11, pax13, pax15, dot16, cho16, spa13}. Assuming the directly imaged brown dwarf companions HD~4747~B and HD~19467~B have the same ages as their respective host stars, we use the isochronal age estimates to test and constrain several substellar evolutionary models (\S\ref{sec:comparisons}) \citep{cha00, bar02, bar03, bar15, sau08}. Both benchmark brown dwarfs have precisely measured dynamical masses and metallicities, making them ideal objects to calibrate models.



\section{Interferometric Observations and Stellar Radii}\label{sec:observations}
In order to obtain direct estimates for the stellar diameters, we performed interferometric observations with Georgia State University's CHARA Array, a long-baseline optical/infrared interferometer located within the Mount Wilson Observatory in California. The CHARA Array consists of six 1-m diameter telescopes with distances between telescopes ranging from $\sim 30 - 330$~meters \citep{ten05}. The predicted angular sizes of HD~4747~A and HD~19467~A, based on the surface brightness relations in \cite{boy14}, are on the order of a few tenths of a milli-arcsecond (mas). Thus, we conducted our observations using the PAVO beam combiner \citep{ire08} in the $R$-band with the baseline configurations listed in Table~\ref{tab:observations} in order to adequately resolve the stars.

\begin{deluxetable*}{rccc}
\tablecaption{Observation Log\label{tab:observations}}
\tablehead{\colhead{\textbf{Object}} & \colhead{\textbf{UT Date}} & \colhead{\textbf{CHARA Baseline}} & \colhead{\textbf{Calibrator}}}
\startdata
HD 4747	&	2015/08/14	&	W1-E1	(313.53 m) &	HD~4622		\\
		&	2016/08/01	&	W2-E2 (156.27 m) &	HD~4622		\\
		&	2016/11/11	&	W1-E2 (251.34 m)	& HD~2696, HD~4622	\\  \\
HD 19467 & 	2014/09/06	& E1-S1 (330.66 m) &	HD~17943, HD~22243	\\
		& 	2014/09/07	& W1-E1	(313.53 m) &	HD~17943, HD~22243	\\
		& 	2015/08/17	& E2-S1 (278.76 m)	&	HD~17943, HD~22243	\\
		& 	2016/11/11	& W1-E2 (251.34 m)	&	HD~16141, HD~17943, HD~22243
\enddata
\tablecomments{Refer to \S\ref{sec:observations} for details.}
\end{deluxetable*}

HD~4747~A was observed during the nights of 14 August 2015 UT, 1 August 2016 UT, and 11 November 2016 UT.
HD~19467~A was observed during the nights of 6 and 7 September 2014 UT, 17 August 2015 UT, and 11 November 2016 UT.
The observations of our targets are bracketed in time with several calibrator stars, the selection of which is based on the JMMC Stellar Diameters Catalog \cite[JSDC;][]{duv16}\footnote{\href{http://www.jmmc.fr/jsdc}{http://www.jmmc.fr/jsdc}.}. In order to identify and thus avoid unknown systematic errors in our interferometry data, we require the use of at least two calibrator stars per target, the use of at least two combinations of telescopes (baselines), and data from at least two nights. Calibrator stars for HD~4747~A are HD~2696 ($\theta_{\rm UD, R} = 0.34\pm0.03$ mas) and HD~4622 ($\theta_{\rm UD, R} = 0.219\pm0.006$ mas). Calibrators for HD~19467~A are HD~16141 ($\theta_{\rm UD, R} = 0.366\pm0.010$ mas), HD~17943 ($\theta_{\rm UD, R} = 0.234\pm0.007$ mas), and HD~22243 ($\theta_{\rm UD, R} = 0.185\pm0.005$ mas) \citep{duv16, che16}.
These calibrators are selected based upon their physical attributes: no known multiplicity, low projected rotational velocity, similar brightness as the respective target in $R$, close angular proximity (max 10 degrees) to the respective science target, and to be unresolved sources based on their estimated angular sizes \citep{vanb05, boy13, von14}. A summary of our observations is shown in Table~\ref{tab:observations}.

\begin{figure*}
\epsscale{1.10}
\plottwo{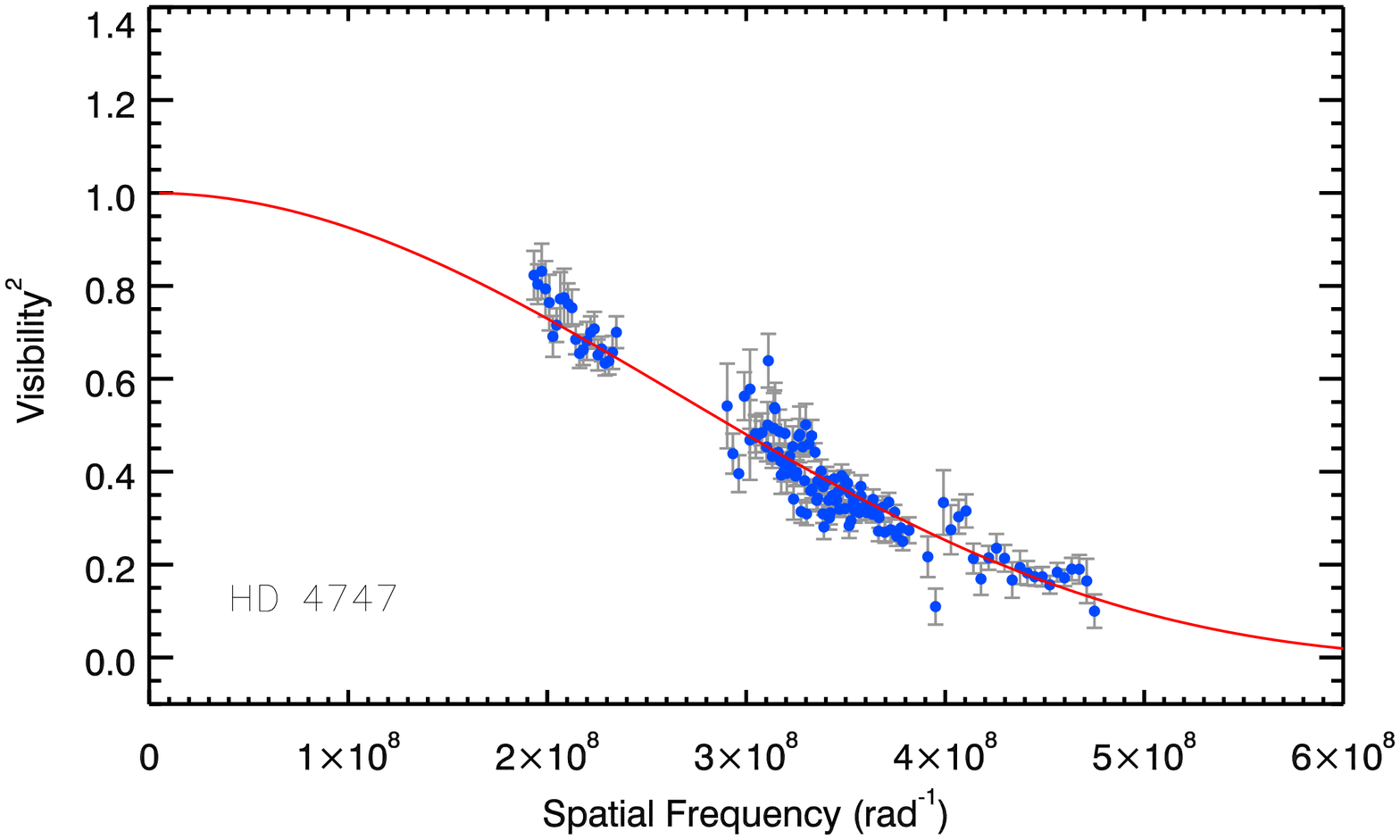}{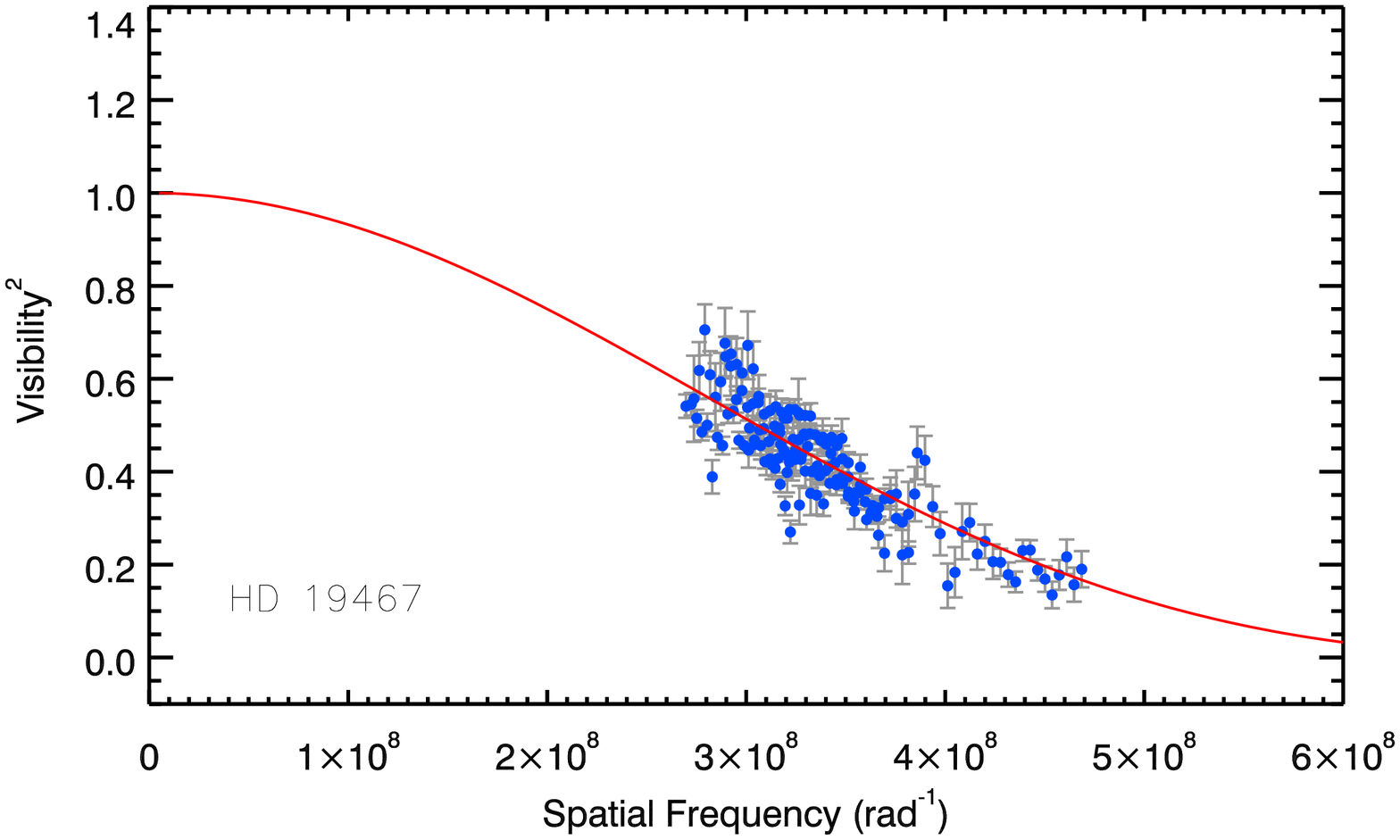}
\caption{\label{fig:visibilities1}Calibrated interferometric $V^2$ values (points) and the $R$-band, limb-darkened fits to those measurements (line) for HD~4747 (left) and HD~19467 (right). The units of the x-axis correspond to the baseline length in unit of operational wavelength. For more details, see \S\ref{sec:observations}.}
\end{figure*}

Our data reduction procedure to extract calibrated squared-visibility measurements ($V^2$, Figure~ \ref{fig:visibilities1}) is described in section 2.1 in \cite{boy15} and is based on the methods outlined in \cite{mae13} and \cite{whi13}. We measure uniform disk angular diameters of $\theta_{\rm UD} = 0.367\pm0.006$ mas for HD~4747~A and $\theta_{\rm UD} = 0.355\pm0.011$ mas for HD~19467~A.
We determine limb-darkened angular diameters of $\theta_{\rm LD} = 0.390\pm0.007$ mas for HD~4747~A and $\theta_{\rm LD} = 0.376\pm0.014$ mas for HD~19467~A using respective limb-darkening coefficients of $\mu_{R} = 0.63$ and $\mu_{R} = 0.60$ \citep{cla11}. Combined with parallaxes from Gaia DR2 \citep{gaia18}, we obtain stellar radii of R = $0.789\pm0.014$ R$_{\sun}$ for HD~4747~A and R = $1.295\pm0.048$ R$_{\sun}$ for HD~19467~A (Table~\ref{tab:propstars}).
Our new radius measurements are consistent with literature values within 1$\sigma$ for HD~4747~A \citep{cre18} and within 3$\sigma$ for HD~19467~A \citep{cre14}.

\begin{deluxetable}{llcc}[hb!]
\tablecaption{\label{tab:propstars}Properties of the Host Stars}
\tablehead{\colhead{Property} & \colhead{} & \colhead{HD~4747~A} & \colhead{HD~19467~A}}
\startdata
RA (J2000)               &  & 00 49 26.77     & 03 07 18.57     \\
Dec (J2000)              &  & -23 12 44.93    & -13 45 42.42    \\
Spectral Type            &  & G9V\tablenotemark{a} & G3V\tablenotemark{b} \\
Parallax (mas)\tablenotemark{c}			  &  & $53.184\pm0.126$ & $31.225\pm0.041$	\\
Distance (pc)           				  &  & $18.80\pm0.04$  & $32.02\pm0.04$  \\
Mass ($M_{\sun}$) &  & $0.82\pm0.04$\tablenotemark{a} & $0.95\pm0.02$\tablenotemark{b} \\
\lbrack Fe/H \rbrack & & $-0.22\pm0.04$\tablenotemark{a} & $-0.15\pm0.02$\tablenotemark{b} \\
log(g) (cm $\rm s^{-2}$) &  & $4.65\pm0.06$\tablenotemark{a} & $4.40\pm0.06$\tablenotemark{b} \\ \hline
$\theta_{\rm UD}$ (mas)\tablenotemark{d}      &  & $0.367\pm0.006$ & $0.355\pm0.011$ \\
$\theta_{\rm LD}$ (mas)\tablenotemark{d}      &  & $0.390\pm0.007$ & $0.376\pm0.014$ \\
Radius ($R_{\sun}$)\tablenotemark{d}      &  & $0.789\pm0.014$ & $1.295\pm0.048$ \\
$F_{\rm BOL}$ ($10^{-8}$~erg~s$^{-1}$~cm$^{-2}$)\tablenotemark{d} &  & $4.02\pm0.03$ & $4.54\pm0.03$ \\
Luminosity ($L_{\sun}$)\tablenotemark{d}  &  & $0.444\pm0.004$ & $1.456\pm0.010$ \\
$T_{\rm eff,\ interferometric}$ (K)\tablenotemark{d}        &  & $5308\pm48$ & $5572\pm104$	 \\
$T_{\rm eff,\ spectroscopic}$ (K)\tablenotemark{d,e}      &  & $5305\pm25$ & $5748\pm25$	 \\
\enddata
\tablenotetext{a}{ \cite{cre16}}
\tablenotetext{b}{ \cite{cre14}}
\tablenotetext{c}{ \cite{gaia18}}
\tablenotetext{d}{ This paper (\S\ref{sec:observations}, \ref{sec:fbol}, \ref{sec:ages})}
\tablenotetext{e}{ Statistical uncertainty only, does not include model uncertainty.}
\end{deluxetable}


\section{Bolometric Fluxes, Stellar Effective Temperatures, and Stellar Luminosities}\label{sec:fbol}
\begin{figure*}
\epsscale{1.10}
\plottwo{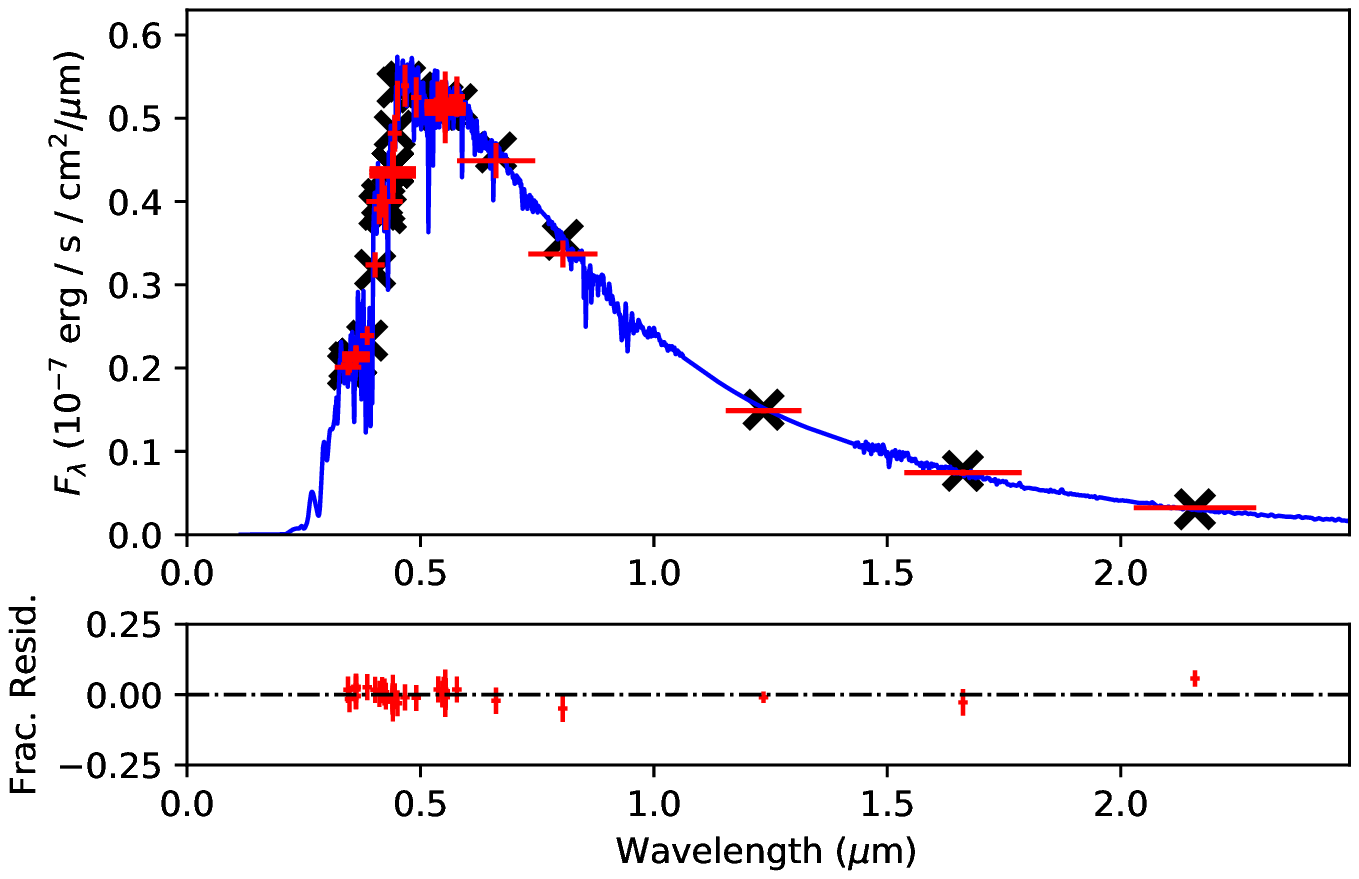}{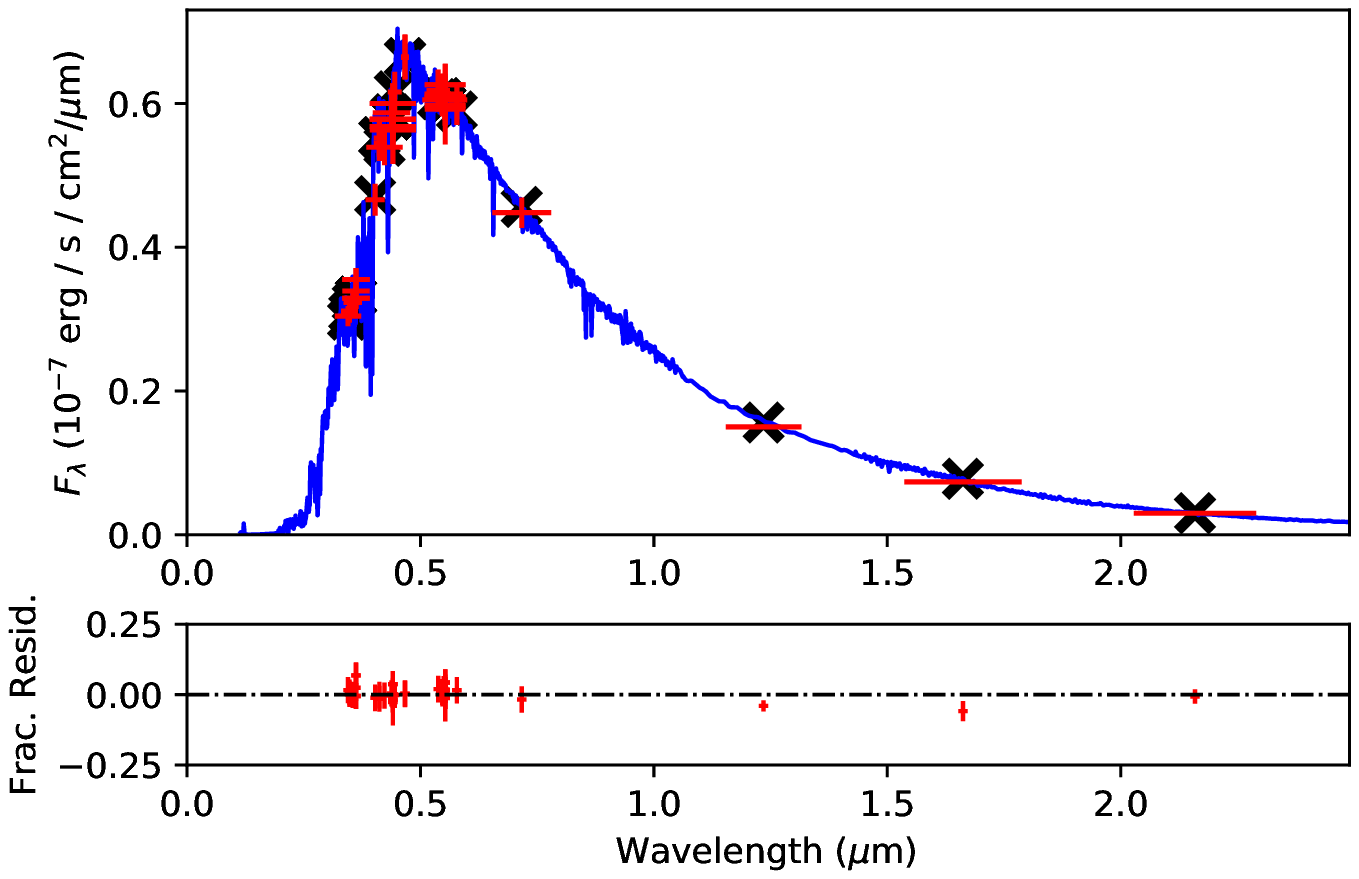}
\caption{\label{fig:seds}SED fits for HD~4747~A (left) and HD~19467~A (right). \cite{pic98} spectral templates (blue lines; G8V for HD~4747~A, G2V for HD~19467~A) are scaled to the literature photometry (red crosses). Black crosses show the flux value of the spectral template integrated over the filter transmission profile. The lower panels display residuals between literature photometry and the spectral template. See \S\ref{sec:fbol} for more details.}
\end{figure*}

Coupled with stellar angular diameter, the knowledge of stellar bolometric flux ($F_{\rm BOL}$) provides a direct estimate of stellar temperature, which, when combined with physical stellar radius, yields stellar luminosity via a reformulation of the Stefan-Boltzmann Law,
\begin{equation} \label{eq:temperature}
T_{\rm eff} ({\rm K}) = 2341 (F_{\rm BOL}/\theta_{\rm LD}^2)^{\frac{1}{4}},
\end{equation}
where $F_{\rm BOL}$ has units of $10^{-8}\ \rm erg/cm^{2}/s$ and $\theta_{\rm LD}$ has units of milliarcseconds.
$F_{\rm BOL}$ can be obtained by spectral energy distribution (SED) fitting by scaling spectral templates to literature photometry values. For the SED fitting of our targets (Figure~ \ref{fig:seds}), we follow the approach used in \cite{man13} and \cite{von14}. Interstellar extinction is set to zero for both targets due to the small distances to our targets (less than 70~pc)\footnote{See \cite{aum09} for more details.} and we use the updated broad-band filter profiles presented in \cite{man15}. In the calculation of the errors in effective temperature and stellar luminosity, we inflate the calculated uncertainty in our $F_{\rm BOL}$ (as given below) by adding 2\% of the error in quadrature, thereby compensating for unknown systematic errors in the literature photometry \citep{boh14}.

Based on fitting a G8V spectral template from the \cite{pic98}\footnote{See also {\href{https://lco.global/~apickles/INGS/}{https://lco.global/$\sim$apickles/INGS/}} for updated spectral templates.} library to literature photometry from \cite{irw61}, \cite{sto63}, \cite{wil69}, \cite{mer86}, \cite{ruf88}, \cite{mer89},
\cite{ols93}, \cite{hau98}, \cite{cut03}, and \cite{koe10}, we measure HD~4747~A's $F_{\rm BOL}$ to be $(4.02\pm0.03) \times 10^{-8}$~erg~s$^{-1}$~cm$^{-2}$, which, when combined with angular diameter as stated in Equation~\ref{eq:temperature},
produces $T_{\rm eff} = 5308\pm48$K and a luminosity of $L = 0.444\pm0.004 L_{\sun}$. Compared to previous literature values, our new temperature estimate for HD~4747~A is consistent within 1$\sigma$ \citep{cre18}.

Using the same approach, we fit a G2V spectral template from the \cite{pic98} library to literature photometry from \cite{cor71}, \cite{cor72}, \cite{ols83}, \cite{egg83}, \cite{mer86}, \cite{ruf88}, \cite{ols94}, \cite{kor96}, \cite{hau98},
and \cite{cut03} to obtain HD~19467~A's $F_{\rm BOL}$ to be $(4.54\pm0.03) \times 10^{-8}$~erg~s$^{-1}$~cm$^{-2}$.
Based on the stellar angular diameter, this yields $T_{\rm eff} = 5573\pm104$K (Equation~\ref{eq:temperature}) and a luminosity of $L = 1.456\pm0.010 L_{\sun}$. Compared with previous literature values, our new temperature estimate for HD~19467~A is consistent within 2$\sigma$ \citep{cre14}.


\section{Stellar Age Estimates}\label{sec:ages}
We derived age estimates for HD~4747~A and HD~19467~A using three different sets of isochrones and two different interpolation procedures. For each estimate, we started with stellar parameters derived from high resolution (R $\sim$ 70,000) spectra of the two stars from the Keck HIRES spectrograph \citep{vog94}, analyzed using the procedure in \citet{bre16}. The procedure uses forward modeling of 350 \AA\ of the spectrum, first fitting for global parameters and deviations from a solar abundance pattern. It then fits for the abundances of 15 elements and repeats the entire procedure using this new abundance pattern. This method has been shown to recover surface gravities consistent with asteroseismology to within $\pm0.05$ dex \citep{bre15}. The effective temperatures obtained from the spectral fitting were consistent with those from the interferometric measurements (Table~\ref{tab:propstars}).

\subsection{Yonsei-Yale Isochrones}\label{sec:yy}
With estimates for [Fe/H], [Si/H] (as a proxy for $\alpha$-element enhancement), $T_{\mathrm{eff}}$, and bolometric luminosity we used the interpolation routines for the YY isochrones from \citet{bre16} to derive masses, radii, surface gravities and ages. The interpolation procedure does not allow us to utilize all of the constraints at our disposal, but the returned radii and surface gravities were consistent with our measured values. One constraint used that is not available for the other interpolation scheme is the Si/Fe ratio. \citet{dot16} showed that stars near their main-sequence turn-off will show an overall depletion of heavy elements in their atmospheres due to diffusion. The ratios of elements remain largely unchanged and so inclusion of this ratio may better capture the abundance of older main sequence stars.

\subsection{MIST and Dartmouth Isochrones}\label{sec:mistdart}
\begin{figure*}[t!]
\plotone{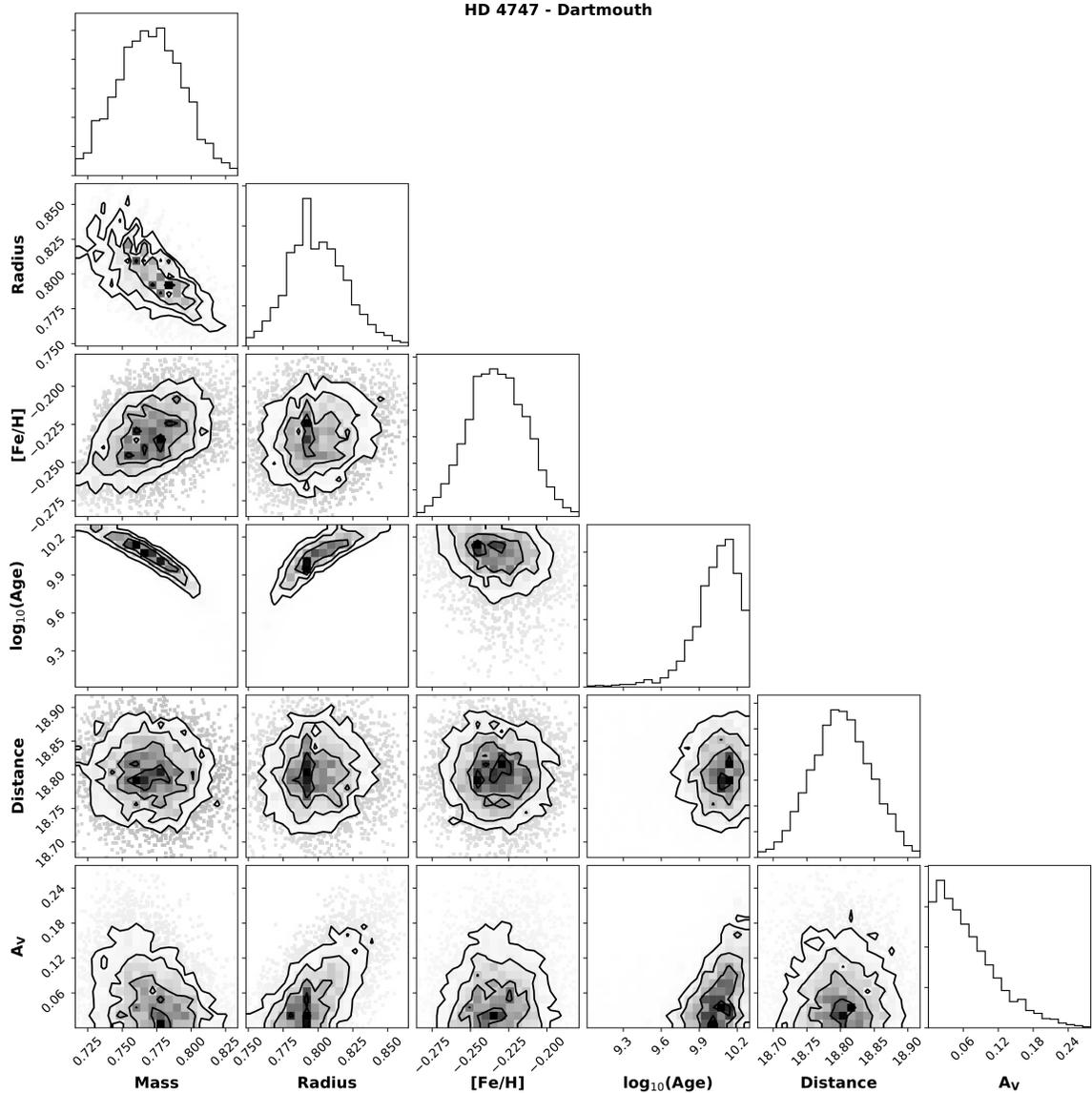}
\caption{\label{fig:dartcorner4747}Corner plot for HD~4747~A from fitting the Dartmouth isochrones. The variables are (from left to right/top to bottom) mass, radius, [Fe/H], $\log_{10}{(age)}$, and $A_{V}$ of the star. The equivalent corner plot for HD~~19467~A and corner plots for both stars from fitting the MIST isochrones are available in the Figure Set. See \S\ref{sec:ages} for more information.}
\end{figure*}

The {\tt isochrones} package \citep{mor15} uses the MultiNest algorithm \citep{fer08, fer09, fer13} to interpolate in either the MIST or Dartmouth isochrone grids. The routine allows for simultaneous fitting of many parameters, which we made use of to include additional constraints not possible with the YY isochrones. For both model grids, we fit the stars using our $T_{\mathrm{eff}}$, $\log g$, [Fe/H], radius, parallax from Gaia DR2 \citep{gaia18}, and V magnitudes. The results of the fitting and correlations can be seen in the corner plots in Figure~\ref{fig:dartcorner4747} and the Figure Set.

\figsetstart
\figsetnum{3}
\figsettitle{Corner Plots for HD~4747~A and HD~19467~A from Isochrone Fitting}

\figsetgrpstart
\figsetgrpnum{3.1}
\figsetgrptitle{HD~4747~A -- Dartmouth}
\figsetplot{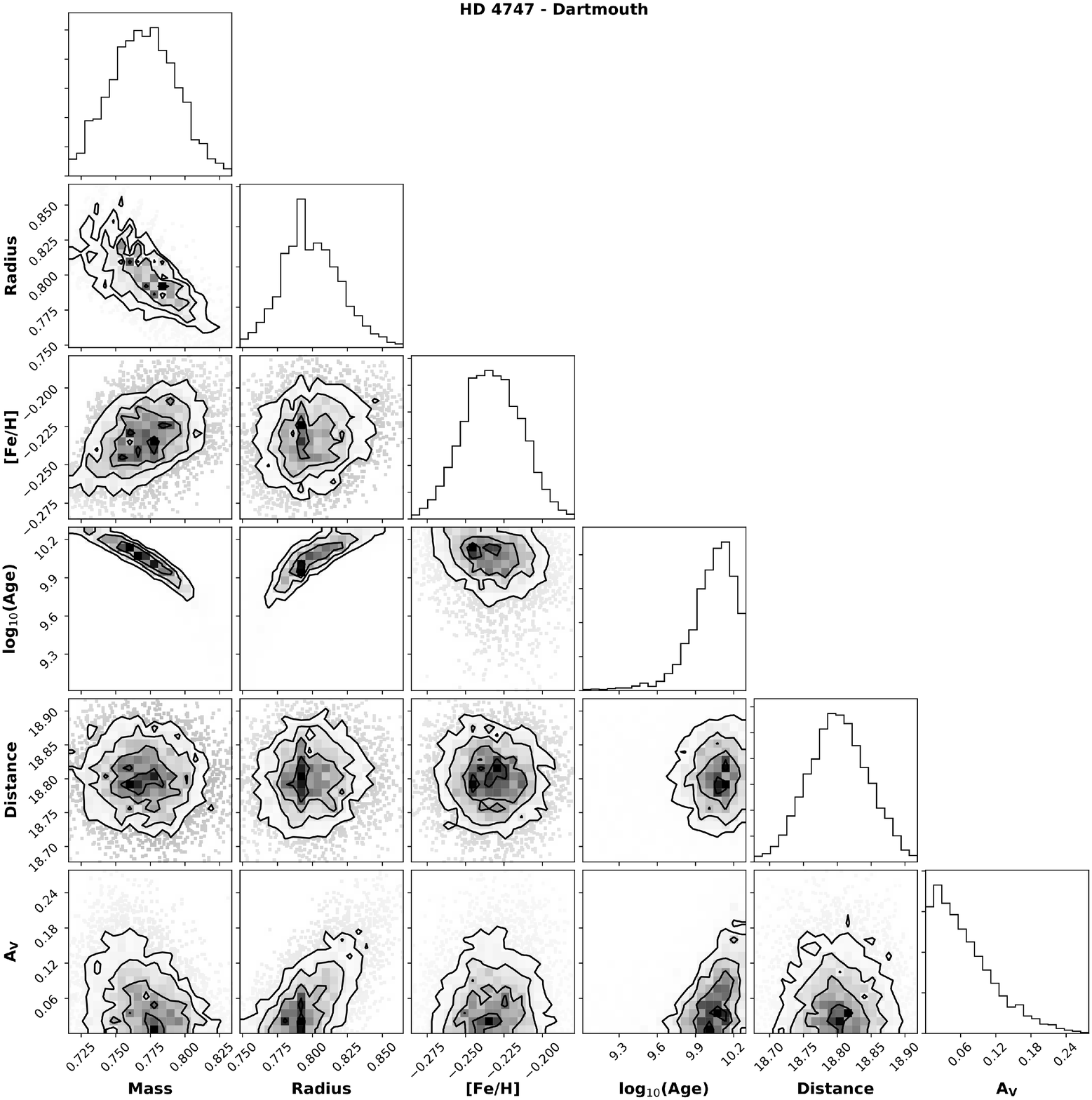}
\figsetgrpnote{Corner plot for HD~4747~A from fitting the Dartmouth isochrones.}
\figsetgrpend

\figsetgrpstart
\figsetgrpnum{3.2}
\figsetgrptitle{HD~4747~A -- MIST}
\figsetplot{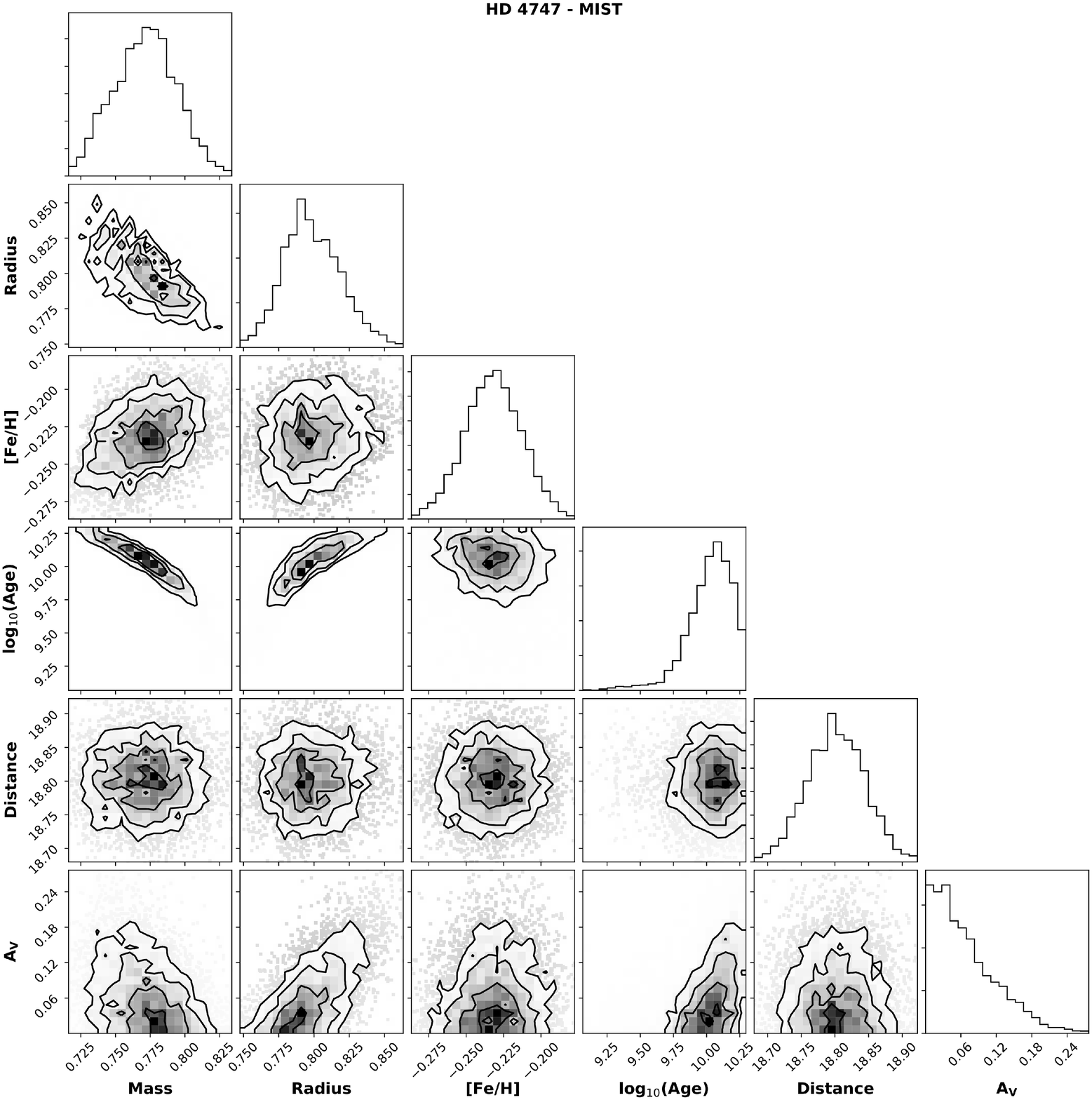}
\figsetgrpnote{Corner plot for HD~4747~A from fitting the MIST isochrones.}
\figsetgrpend

\figsetgrpstart
\figsetgrpnum{3.3}
\figsetgrptitle{HD~19467~A -- Dartmouth}
\figsetplot{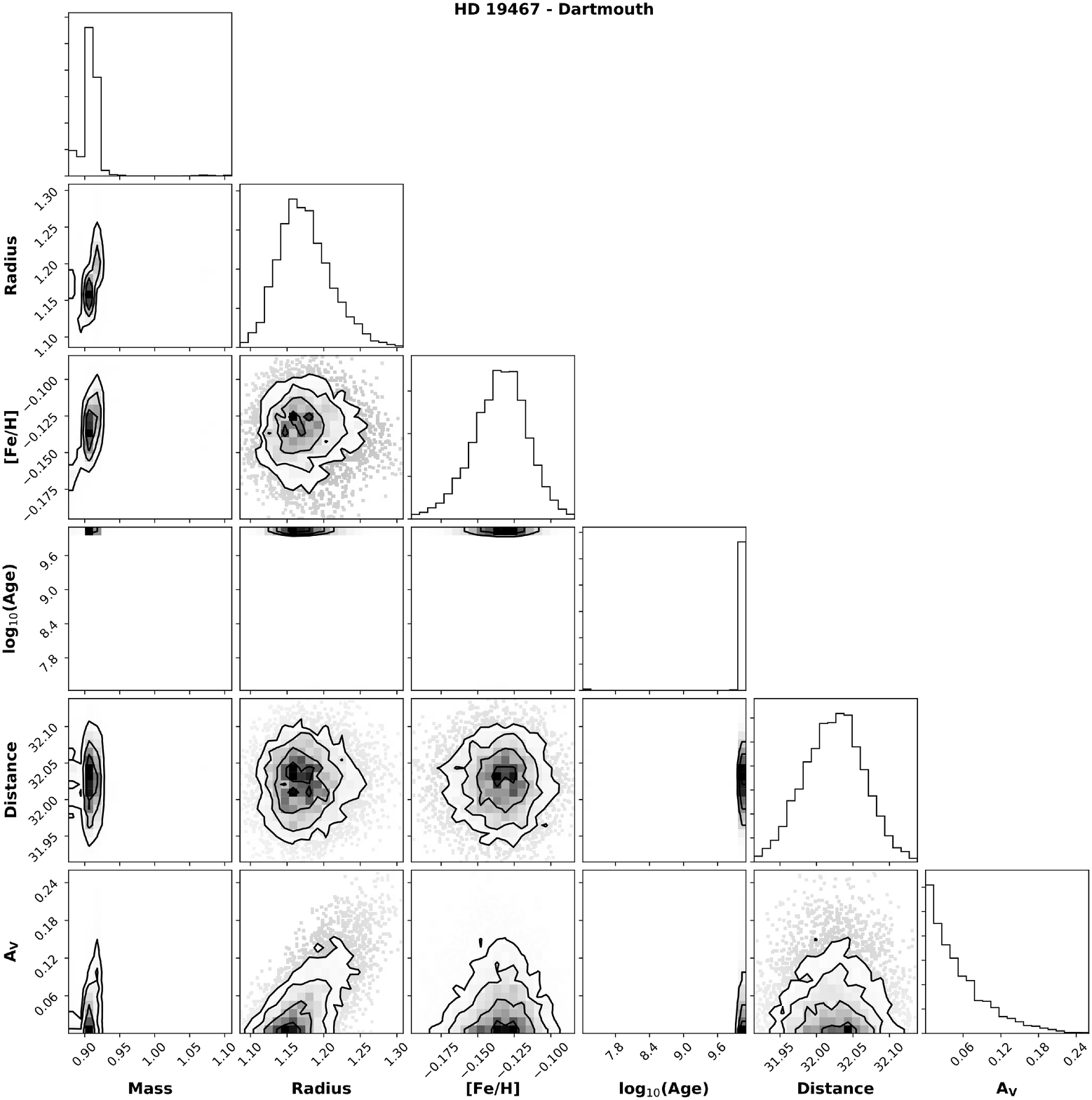}
\figsetgrpnote{Corner plot for HD~19467~A from fitting the Dartmouth isochrones.}
\figsetgrpend

\figsetgrpstart
\figsetgrpnum{3.4}
\figsetgrptitle{HD~19467~A -- MIST}
\figsetplot{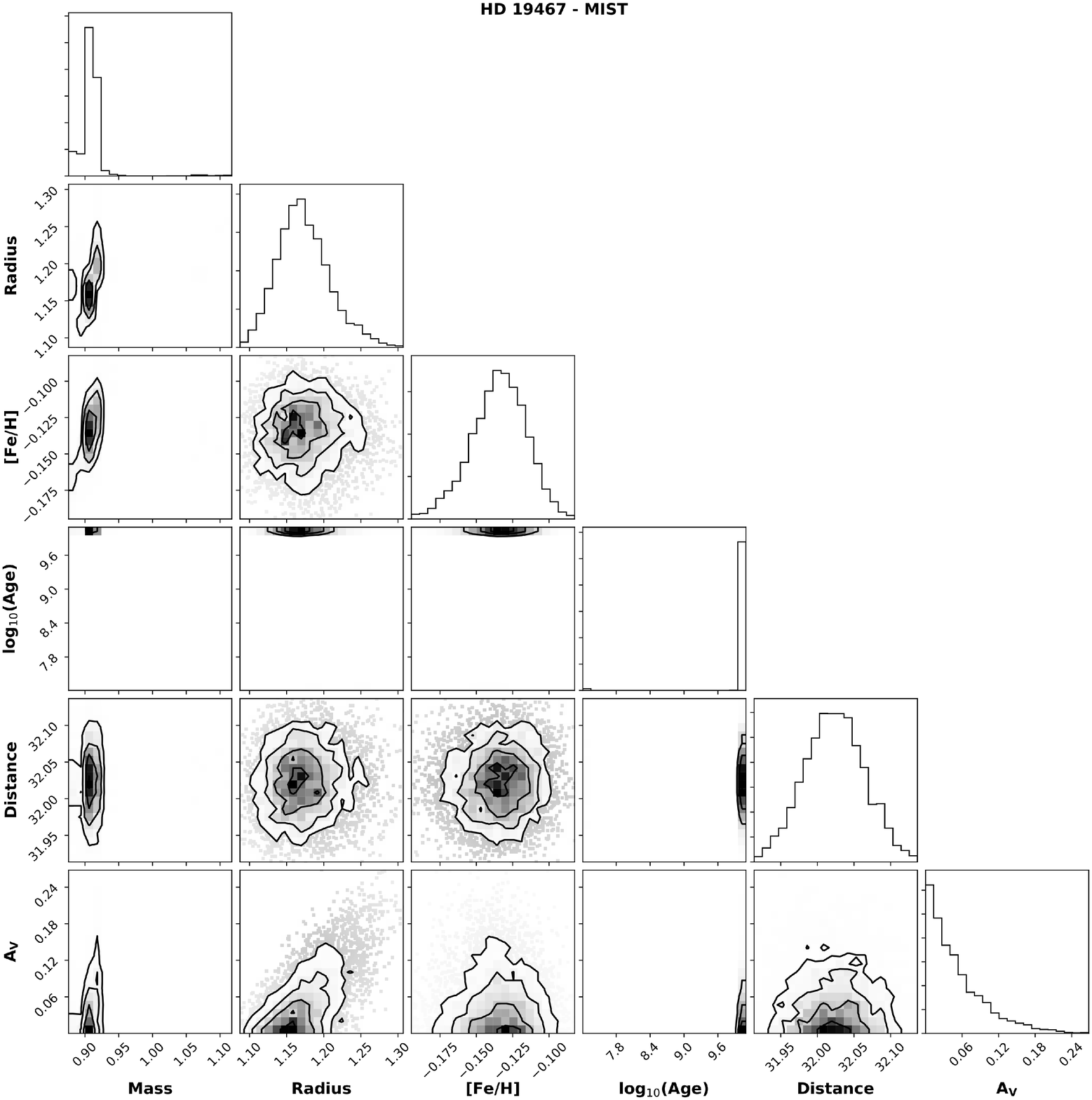}
\figsetgrpnote{Corner plot for HD~19467~A from fitting the MIST isochrones.}
\figsetgrpend

\figsetend

\subsection{Isochrone Age Results}\label{sec:ageresults}

\begin{deluxetable}{llcc}
\tablecaption{\label{tab:ages}Summary of Isochronal Age Estimates (Gyr) for HD~4747~A and HD~19467~A}
\tablehead{\colhead{Isochrone} & \colhead{} & \colhead{HD~4747~A} & \colhead{HD~19467~A}}
\startdata
Dartmouth     &  & $11.33^{+4.37}_{-4.25}$ & $10.66\pm0.51$ \\
MIST      		&  & $11.49^{+4.25}_{-4.27}$ & $10.66^{+0.50}_{-0.51}$ \\
Yonsei-Yale 	&  & $9.39^{+2.90}_{-3.30}$ & $8.85^{+0.92}_{-0.40}$ \\ \hline
Adopted Age\tablenotemark{a} &  & $10.74^{+6.75}_{-6.87}$ & $10.06^{+1.16}_{-0.82}$ \\
\enddata
\tablenotetext{a}{ Calculated as an average between the three age estimates. See \S\ref{sec:ageresults}.}
\end{deluxetable}

The results for both stars and all three isochrone grids is summarized in Table~\ref{tab:ages}. Ages for HD~4747~A were consistent among the three different isochrone grids, though the uncertainties were large and the YY ages were lower by several Gyrs. Low mass stars on the main sequence spend a large amount of time with only minimal changes in their temperature and brightness, making precise age determinations challenging. The YY age estimate for HD~19467~A was also lower than that for the MIST or Dartmouth estimates, which were again consistent with one-another. In all three cases, HD~19467~A is fit to be on the sub-giant branch and has much smaller age uncertainties due to the rapid evolution in this region.

The low age from the YY isochrones could be due to the inclusion of the Si/Fe ratio and its additional constraint on the initial metallicity. However, the MIST isochrones also place additional constraints on the initial metallicity by using surface abundances instead of initial abundances. Instead, the systematically lower ages from YY for both stars points to a difference in the stellar structure of the models at older evolutionary states, resulting in an age offset.

Unlike the Dartmouth and MIST isochrones, the YY isochrones do not allow us to include the surface gravity as a constraint. Since the surface gravities are consistent to within $\pm0.05$ dex of those from asteroseismology, we trust the Dartmouth and MIST age estimates over the YY age estimates, however we still include the YY estimates in our analysis. We adopt ages that are averages of the Dartmouth, MIST, and YY estimates: $10.74^{+6.75}_{-6.87}$ Gyr for HD~4747~A and $10.06^{+1.16}_{-0.82}$ Gyr for HD~19467~A.

\subsection{Discrepancies Between Age Estimates}\label{sec:discrepancies}
The gyrochronological age estimates for HD~4747~A \cite[$3.3^{+2.3}_{-1.9}$ Gyr;][]{cre16} and HD~19467~A \cite[$4.3^{+1.0}_{-1.2}$ Gyr;][]{cre14} are several Gyr younger than the isochronal age estimates. One possible explanation for this discrepancy is tidal interactions with a nearby companion ``spinning up" the star \citep{bro14, max15}. This seems unlikely, as the only known massive companions to both HD~4747~A and HD~19467~A are the benchmark brown dwarfs separated by $\rho = 11.3\pm0.2$ AU and $\rho = 51.1\pm1.0$ AU respectively. A more probable explanation is weakened magnetic braking, which occurs in solar-type stars with ages $\geq 4-5$ Gyr \citep{vans16}. This would result in gyrochronological age estimates of around 4 Gyr, despite the actual age of the star being older.

Isochronal models become less reliable as a star's properties deviate from those of the Sun \citep{bona12, tay17}. However, both HD~4747~A and HD~19467~A are nearly Sun-like in mass, radius, luminosity, and metallicity, so we expect the isochronal models to be well-calibrated. In addition, gyrochronology is only precisely constrained for stars younger than the Sun \citep{mam08}. As a result, we adopt the isochronal age estimates (Table~\ref{tab:ages}) over the gyrochronological age estimates for both HD~4747~A and HD~19467~A.

To further investigate the discrepancy between isochronal and gyrochronological age estimates, the ages of HD~4747~A and HD~19467~A could be determined using asteroseismology \citep{ulr86, leb14, sil15}. While neither star is on the TESS Asteroseismic Science Consortium (TASC) target list due to lower probabilities of detection of solar-like oscillations \cite[about 5\% for HD~4747~A and 20\% for HD~19467~A;][]{cam16}, it is worth looking at since they are both relatively high on the Candidate Target List and should still be targeted with the two-minute cadence \citep{sta17}. Other methods of determining age that are related to stellar activity or rotation, such as measuring lithium abundance or X-ray emission, would be correlated with the gyrochronological age and therefore not useful for resolving the discrepancy.


\section{Bolometric Luminosities of HD~4747~B and HD~19467~B}\label{sec:bollum}

We calculate the bolometric luminosities of the brown dwarfs following the method outlined in Appendix A of \cite{cre12} using the following equations:
\begin{equation}
M_{bol} = M_{K_s} - 0.11 + BC_{K}
\end{equation}
\begin{equation} \label{eq:luminosity}
L = 10^{(M_{bol,\sun}-M_{bol})/2.5} L_{\sun}
\end{equation}
where the bolometric magnitude of the Sun $M_{bol,\sun} = 4.74$.

HD~4747~B has an absolute magnitude $M_{K_{s}} = 13.00\pm0.14$ \citep{cre16}. Combined with the 0.11 mag correction to convert to $M_{K}$ \citep{rud96} and an estimated bolometric correction $BC_{K} = 2.93\pm0.09$ \citep{gol04} using the updated spectral type and temperature from \cite{cre18}, we obtain a bolometric magnitude $M_{bol} = 15.82\pm0.17$. This gives us a bolometric luminosity $L = (3.70\pm0.57) \times 10^{-5} L_{\sun}$.

HD~19467~B has an absolute magnitude $M_{K_{s}} = 15.52\pm0.10$ \citep{cre14}. Combined with the 0.11 mag correction to convert to $M_{K}$ \citep{rud96} and an estimated bolometric correction $BC_{K} = 2.30\pm0.13$ \citep{gol04}, we obtain a bolometric magnitude $M_{bol} = 17.71\pm0.16$. This gives us a bolometric luminosity $L = (6.49\pm0.98) \times 10^{-6} L_{\sun}$.

\begin{deluxetable}{llcc}[h!]
\tablecaption{\label{tab:propdwarfs}Properties of the Brown Dwarf Companions}
\tablehead{\colhead{Property} & \colhead{} & \colhead{HD~4747~B\tablenotemark{a}} & \colhead{HD~19467~B\tablenotemark{b}}}
\startdata
Spectral Type            &  & T1$\pm$2        & T5 - T7         \\
Separation (AU)          &  & $11.3\pm0.2$    & $51.1\pm1.0$    \\
\lbrack Fe/H \rbrack     &  & $-0.22\pm0.04$  & $-0.15\pm0.02$  \\
$T_{\rm eff}$ (K)        &  & $1450\pm50$     & $1050\pm40$	    \\
Luminosity ($L_{\sun}$)\tablenotemark{c} &  & $3.70\pm0.57 \times 10^{-5}$ & $6.49\pm0.98 \times 10^{-6}$ \\
\enddata
\tablenotetext{a}{ \cite{cre16, cre18}}
\tablenotetext{b}{ \cite{cre14}}
\tablenotetext{c}{ This work (\S\ref{sec:bollum})}
\end{deluxetable}


\section{Comparison to Substellar Evolutionary Models}\label{sec:comparisons}
Assuming the brown dwarf companions HD~4747~B and HD~19467~B have the same ages as their respective host stars, we can directly test the accuracy of several substellar evolutionary models (SSEM). For this paper, we looked at SSEMs from \cite{bar03} (COND03), \cite{bar15} (BHAC15), and \cite{sau08} (SM08) and compared them to calculated properties of HD~4747~B and HD~19467~B both graphically and numerically.

\subsection{Visual Comparisons}\label{sec:viscomp}
\begin{figure*}[hb!]
\gridline{\fig{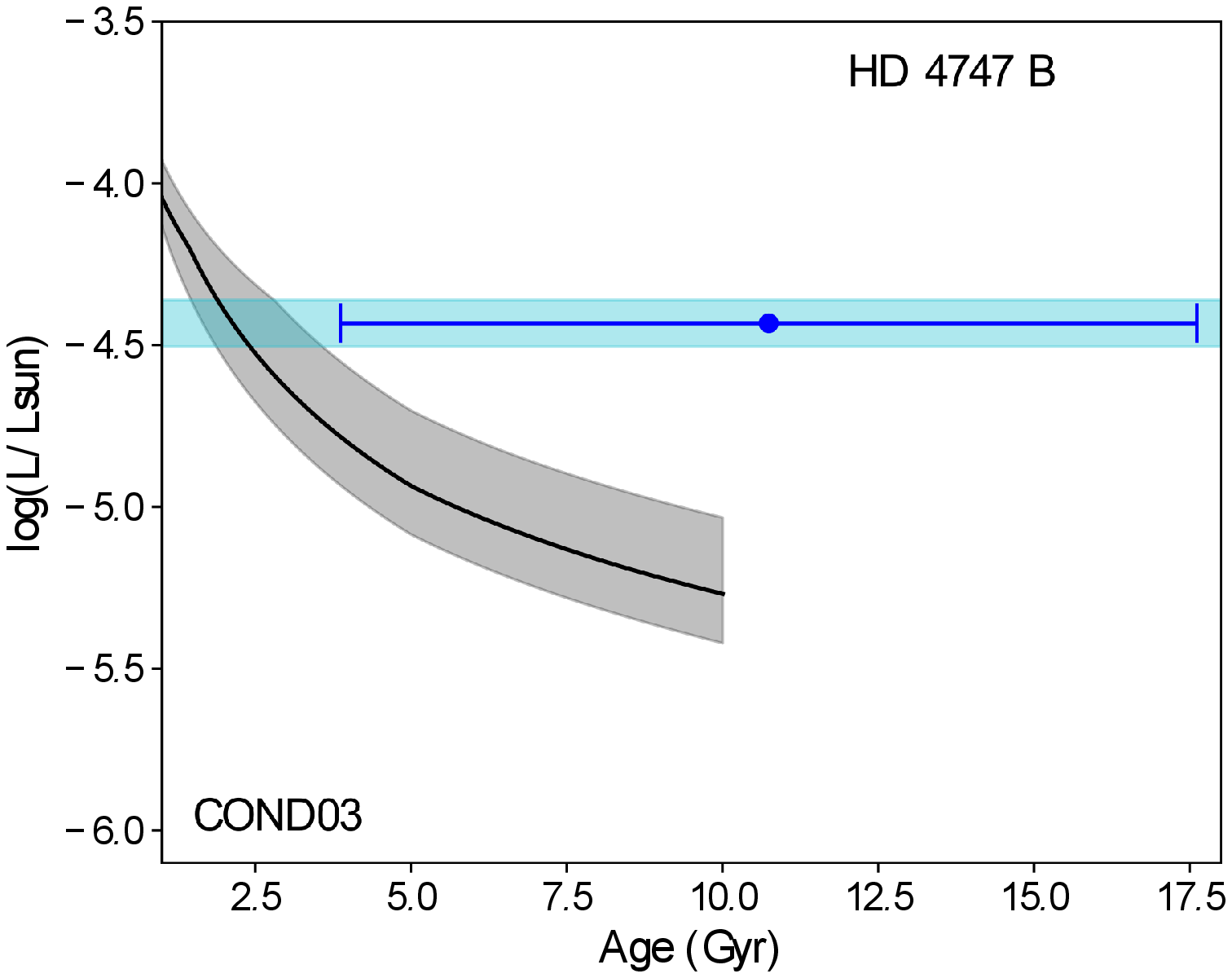}{0.50\textwidth}{(a)}
		  \fig{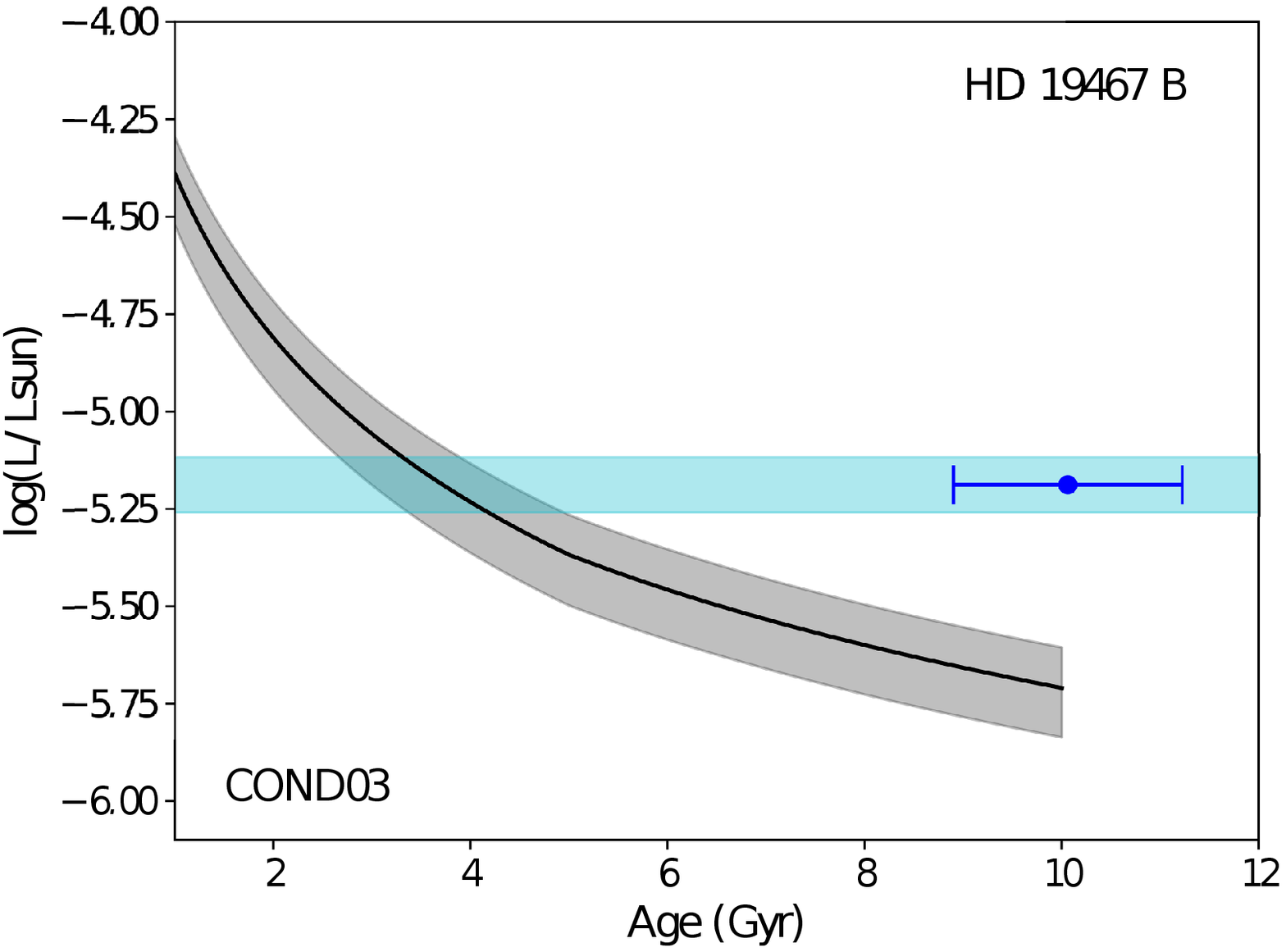}{0.50\textwidth}{(c)}}
\gridline{\fig{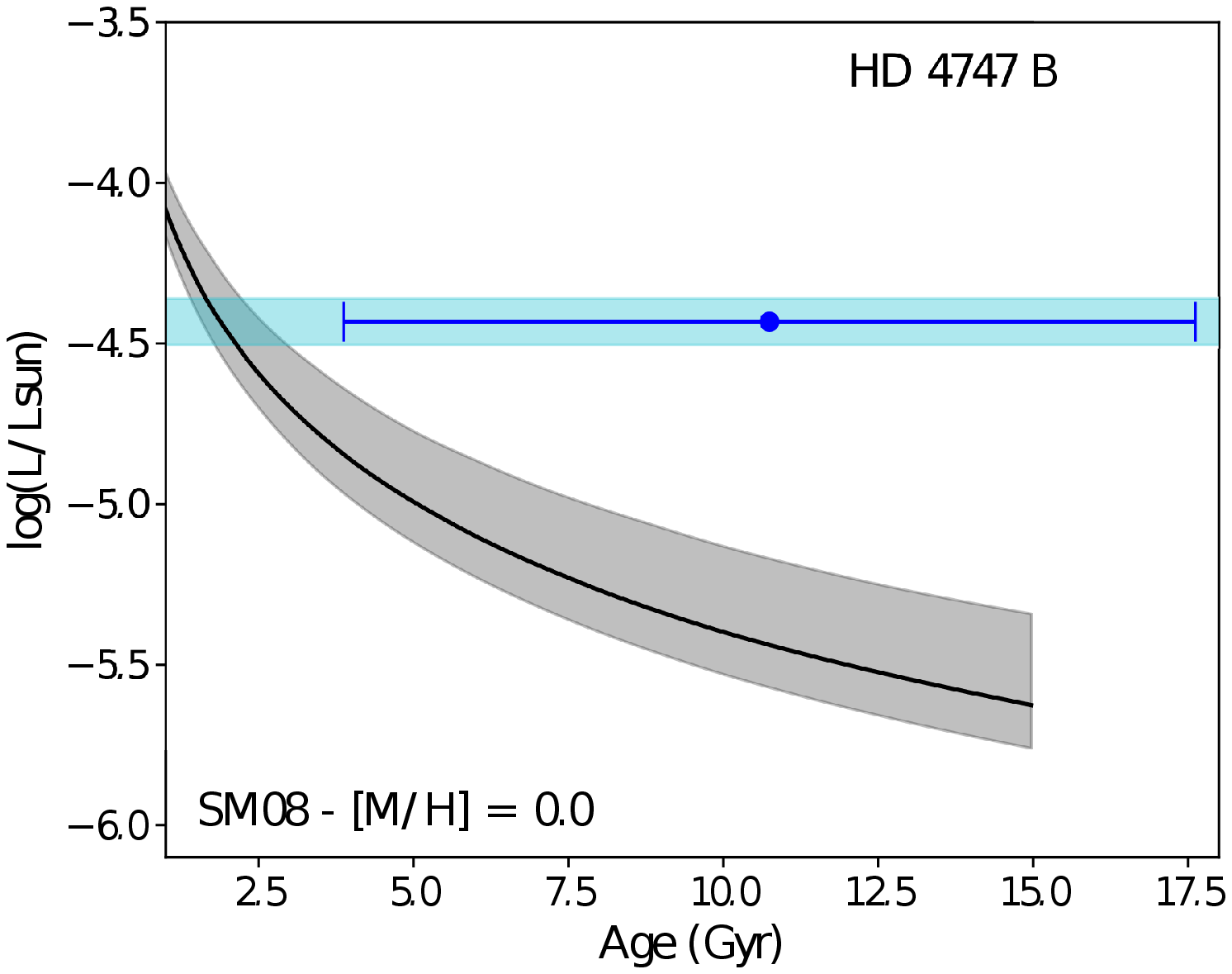}{0.50\textwidth}{(b)}
		  \fig{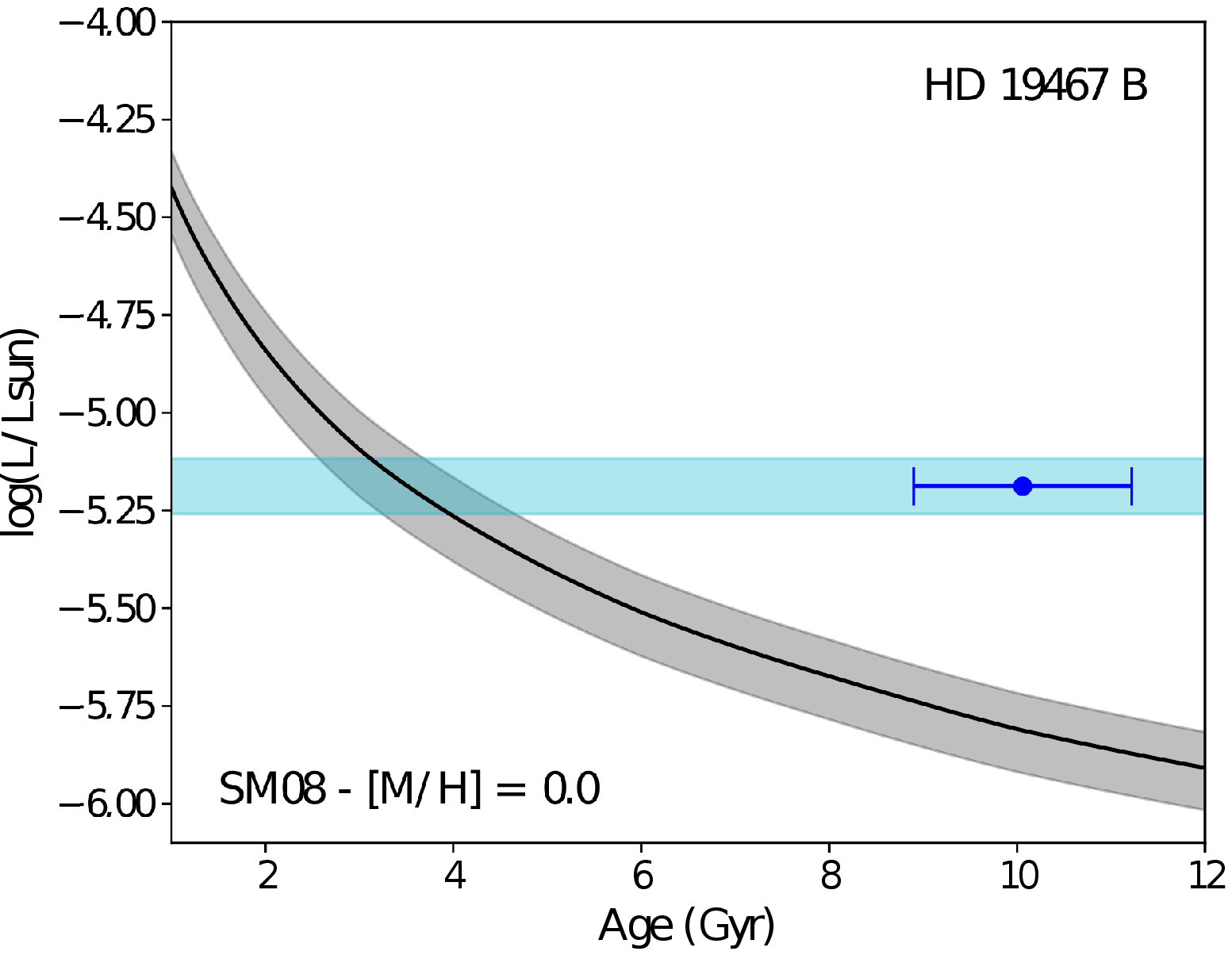}{0.50\textwidth}{(d)}}
\caption{\label{fig:cloudless}Luminosity vs. age comparison of the COND03 and SM08 substellar evolutionary models (black curves) to the observed data (blue dots) for HD~4747~B (a and b) and HD~19467~B (c and d). The light blue bars correspond to the uncertainty in the bolometric luminosities for the brown dwarfs. Although the models do not extend past 10 Gyr (COND03) and 15 Gyr (SM08), it is clear that they under-predict the bolometric luminosities of both objects because brown dwarfs do not sustain fusion and continuously cool. The models are too low by $\sim 0.75$ dex for HD~4747~B and $\sim 0.5$ dex for HD~19467~B. See \S\ref{sec:viscomp}.}
\end{figure*}

We directly compare the calculated bolometric luminosities (\S\ref{sec:bollum}) for the brown dwarfs to the theoretical predictions from each SSEM given their dynamical masses and isochronal age estimates. Each SSEM is linearly interpolated across ages and masses using the SciPy \citep{jon01} algorithm {\tt LinearNDInterpolator} to give a grid of bolometric luminosity predictions, with age and mass spanning ranges determined by the extent of each model\footnote{Ages generally range from 0.0010 to 10 Gyr and masses generally range from 0.001 to 0.072 $M_{\sun}$.}. We then plot the SSEM linear interpolations with the data points for each brown dwarf to see if they are consistent (Figure~\ref{fig:cloudless}).

We find that the COND03 and SM08 SSEMs under-predict the bolometric luminosities of both brown dwarf companions, which is consistent with previous tests of SSEMs using benchmark brown dwarfs \citep{dup09a, cre12, dup14, cre18}. The model predictions are too low by $\sim0.75$ dex for HD~4747~B at the best-fit age and mass and $\sim0.5$ dex for HD~19467~B. If the masses of both objects are slightly higher, which has been suggested for HD~4747~B \citep{per18}, the measured bolometric luminosity would be more consistent with the models. For HD~4747~B, increasing the mass places the object around the hydrogen burning limit, increasing the range in the predicted luminosity. We do not make any conclusions regarding the BHAC15 models at this time as they currently do not extend to ages older than $\sim 2$ Gyr for masses lower than 0.080 $M_{\sun}$.

\subsubsection{Effects of Metallicity on Luminosity Predictions}\label{sec:metallicity}
There are few SSEMs available that explore the effect of metallicity on brown dwarf evolution. Of the models tested, only \cite{sau08} provide grids for metallicities other than solar. To effectively explore how metallicity changes the luminosity predictions of brown dwarfs, SSEMs that span a wider range of metallicities are needed, such as the upcoming Sonora models \citep{mar17}.

\begin{figure*}[ht!]
\gridline{\fig{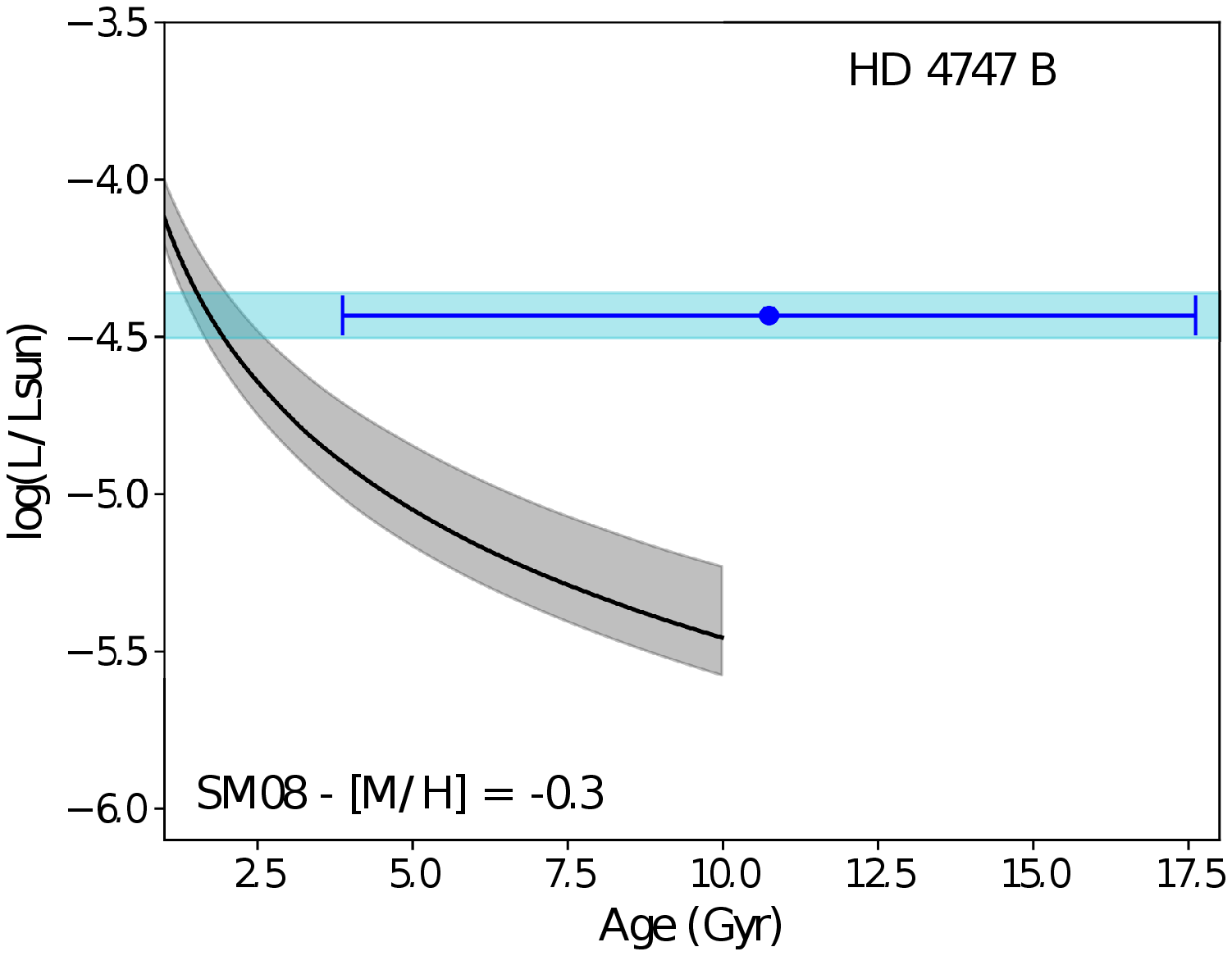}{0.50\textwidth}{(a)}
		  \fig{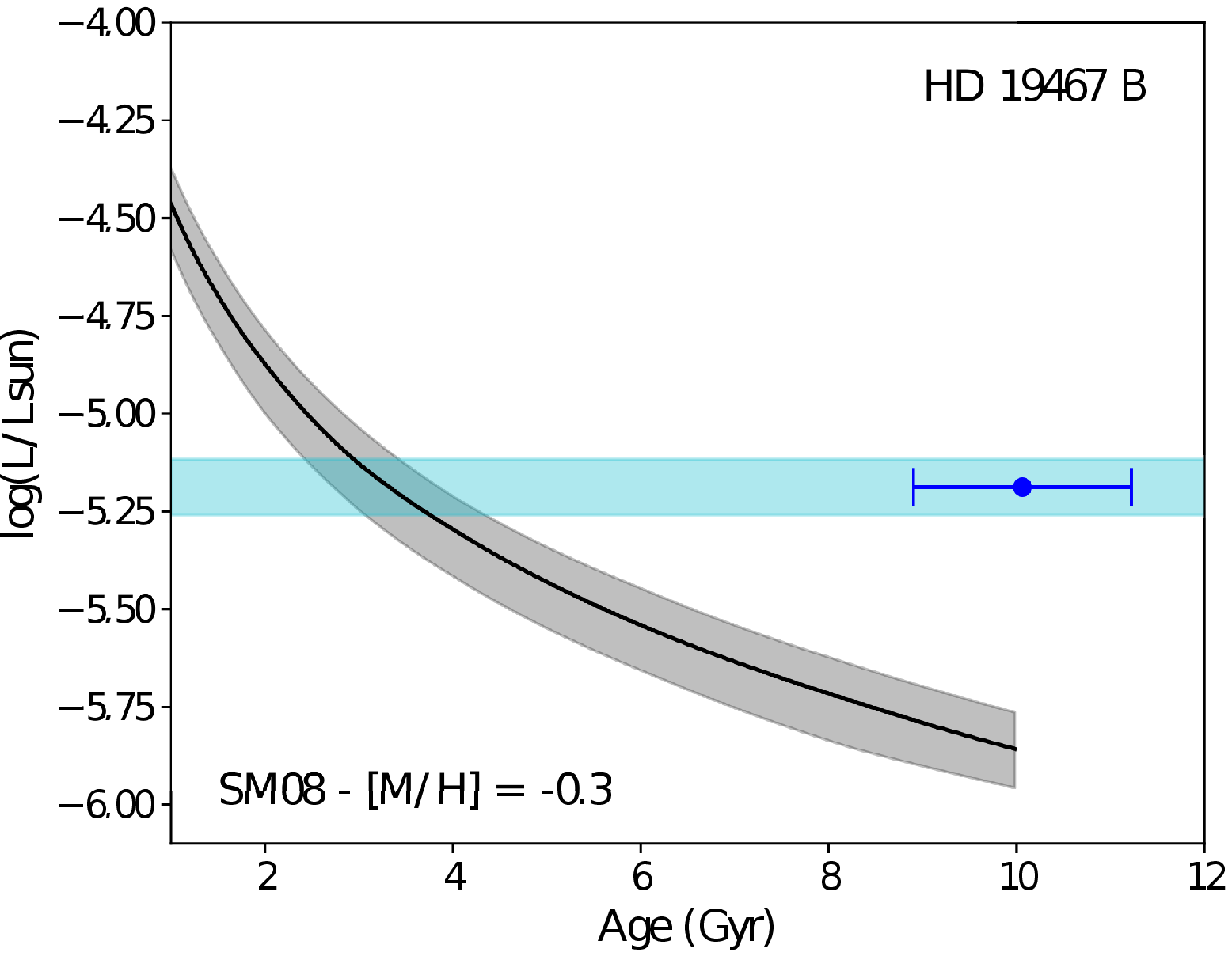}{0.50\textwidth}{(b)}}
\caption{\label{fig:metallicity}Luminosity vs. age comparison of the lower metallicity SM08 substellar evolutionary model (black curves) to the observed data (blue dots) for HD~4747~B (a) and HD~19467~B (b). The light blue bars correspond to the uncertainty in the bolometric luminosities for the brown dwarfs. For both HD~4747~B and HD~19467~B, the lower metallicity model increases the discrepancy to $\sim 1$ dex and $\sim 0.7$ dex respectively. See \S\ref{sec:metallicity}.}
\end{figure*}

 Since both HD~4747~B and HD~19467~B have metallicities slightly less than solar, we compare them to the grid assuming \lbrack M/H \rbrack = -0.3 (Figure~\ref{fig:metallicity}). In both cases, this comparison does not improve the discrepancy between the calculated and predicted bolometric luminosities. The lower metallicity model under-predicts the bolometric luminosity for HD~4747~B by $\sim 1$ dex and $\sim 0.7$ dex for HD~19467~B.

\begin{figure*}[ht!]
\gridline{\fig{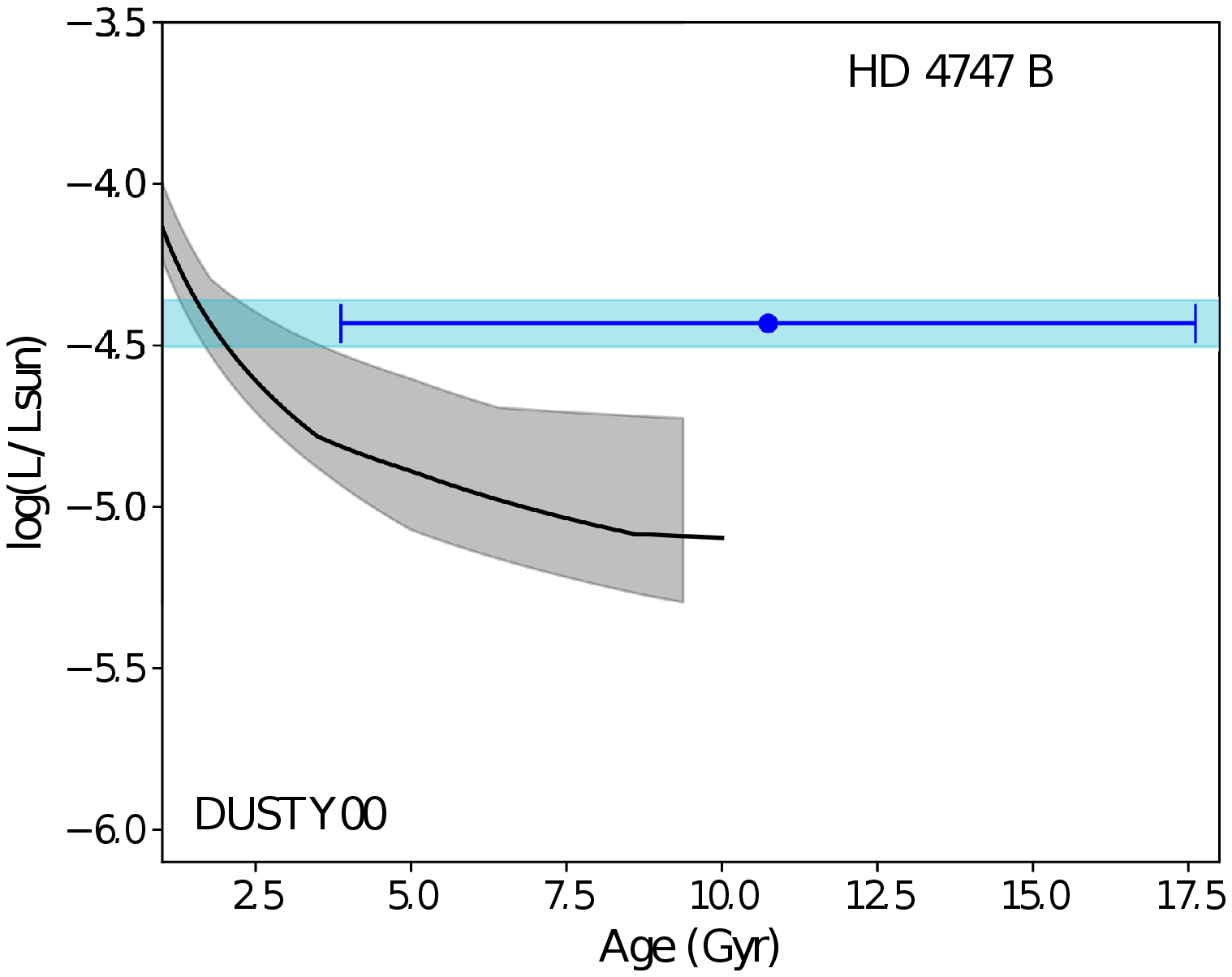}{0.50\textwidth}{(a)}
		  \fig{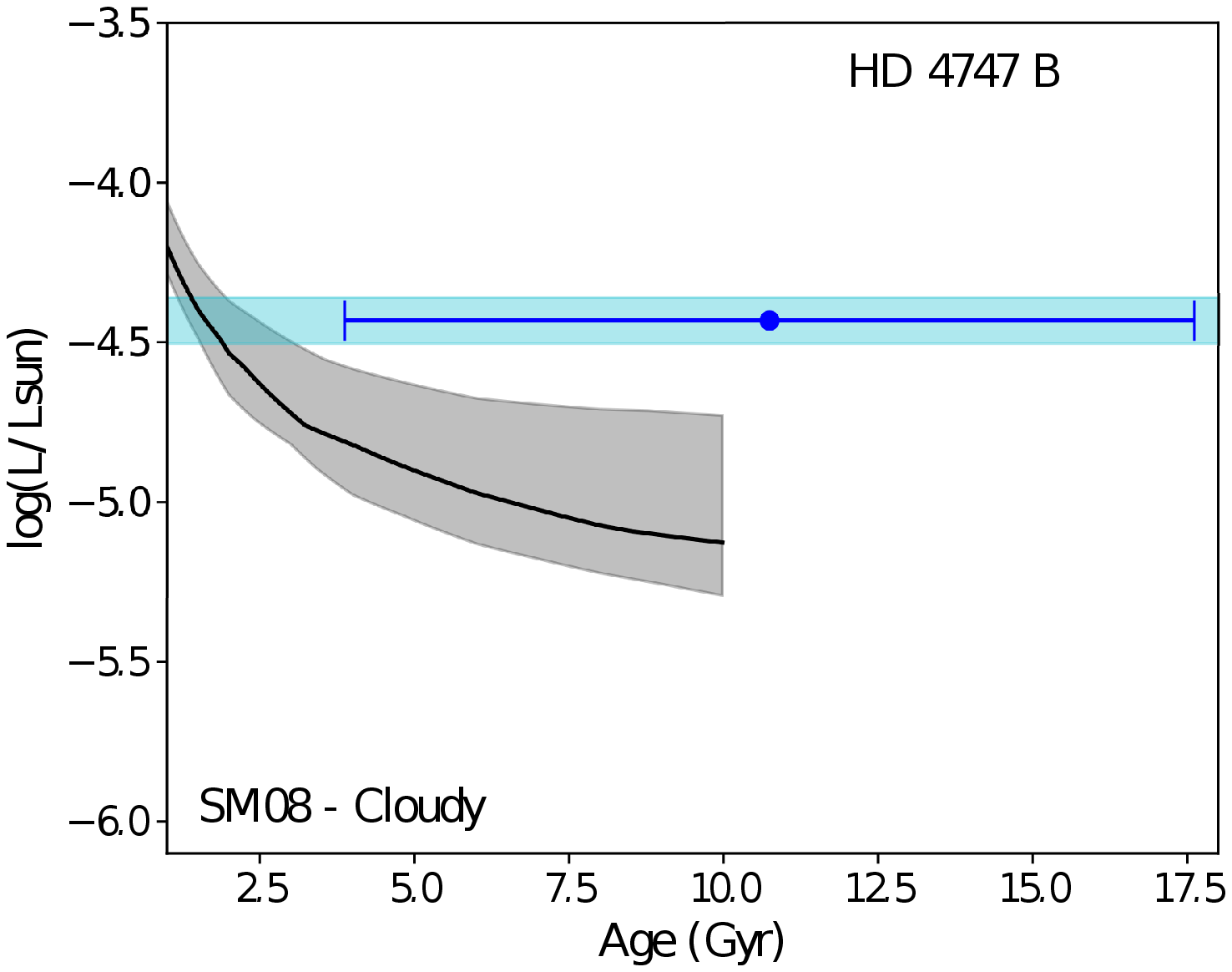}{0.50\textwidth}{(b)}}
\caption{\label{fig:cloudy}Luminosity vs. age comparison of the DUSTY00 and SM08-C substellar evolutionary models (black curves) to the observed data (blue dots) for HD~4747~B. The light blue horizonal bars correspond to the uncertainty in the bolometric luminosity for the brown dwarf. The cloudy models predict the luminosity of HD~4747~B better than the cloudless  models, reducing the discrepancy to $\sim0.6$ dex. See \S\ref{sec:clouds}.}
\end{figure*}

\subsubsection{Effects of Clouds on Luminosity Predictions for HD~4747~B}\label{sec:clouds}
HD~4747~B is an early T-dwarf (spectral type T1$\pm$2) near the L/T transition, where its atmosphere is cool enough to begin forming clouds \citep{cre16, cre18}. To account for this, we also compare HD~4747~B to SSEMs that include cloud formation by \cite{cha00} and \cite{bar02} (DUSTY00) and \cite{sau08} (SM08-C) (Figure~\ref{fig:cloudy}). The cloudy models are a closer fit to the data for HD~4747~B than the cloudless models, reducing the discrepancy in the bolometric luminosity to $\sim 0.6$ dex.


\subsection{Photometric Mass Estimates}\label{sec:numcomp}
Using a Markov-Chain Monte Carlo (MCMC) simulation, we calculate the photometric mass of HD~4747~B and HD~19467~B according to each SSEM given their isochronal ages and bolometric luminosities. We perform the MCMC simulation using the Python package {\tt emcee} \citep{for13}, which implements an affine-invariant ensemble sampler to explore our three-dimensional (age, mass, and luminosity) parameter space with Gaussian priors on age and luminosity and a Gaussian likelihood function for mass. The results of the MCMC are shown in Table~\ref{tab:photomass}.

As expected based on our plots from \S\ref{sec:viscomp}, the predicted photometric masses for both HD~4747~B and HD~19467~B are higher ($\sim 12\%$ and $\sim 30\%$ respectively) than the dynamical mass measurements when considering the cloudless models. For HD~19467~B, the models are discrepant by about $4\sigma$. Due to the larger lower bound errors from the models for HD~4747~B, the cloudless models are consistent with the dynamical mass. When comparing to the cloudy models, the predicted mass for HD~4747~B is reduced to $\sim 8\%$ higher than the dynamical mass, which is still consistent within $1\sigma$.


\begin{deluxetable}{l|cc}[ht!]
\tablecaption{\label{tab:photomass}Photometric Masses of HD~4747~B and HD~19467~B}
\tablehead{\colhead{Mass Model} & \colhead{HD~4747~B} & \colhead{HD~19467~B}}
\startdata
Dynamical\tablenotemark{a} & $65.3^{+4.4}_{-3.3}$ & $51.9^{+3.6}_{-4.3}$ \\ \hline
COND03 & $72.7^{+3.4}_{-13.6}$ & $67.3^{+0.9}_{-1.2}$ \\
SM08 & $74.3^{+1.2}_{-11.5}$ & $68.6^{+1.2}_{-1.6}$ \\
DUST00 & $69.7^{+1.6}_{-13.6}$ & -- \\
SM08-C & $71.7^{+1.2}_{-9.3}$ & -- \\
\enddata
\tablecomments{Masses reported in units of $M_{Jup}$.}
\tablenotemark{a}{ \cite{cre18, cre14}}
\end{deluxetable}

\section{Summary and Conclusions}\label{sec:conclusion}

For brown dwarfs found as companions to stars, certain properties such as metallicity and age can be determined independent from the brown dwarf's mass and luminosity by studying the host star instead of the brown dwarf. As a result, such objects are ideal to use as benchmarks for substellar evolutionary models. While not many are known, benchmark brown dwarfs tend to be over-luminous compared to SSEMs.

Using new age estimates for HD~4747~B and HD~19467~B, determined by studying the host stars with interferometry, we have shown that current SSEMs under-predict the bolometric luminosities and over-predict the masses of these brown dwarfs. Our discrepancy between measured and predicted bolometric luminosities is high compared to previous results for HD~130948~BC and HR~7672~B \citep{dup09a, cre12}, but the discrepancy between measured and predicted masses is consistent with results for Gl~417~BC \citep{dup14}. Since both HD~4747~B and HD~19467~B orbit far from their host stars, we do not expect this additional luminosity to result from heating due to the star.

Although including clouds in the SSEMs puts the predicted mass and luminosity of HD~4747~B in better agreement to the measured data, the brown dwarf still appears over-luminous. A possible explanation for the remaining discrepancy is missing physics in the models. The effect of metallicity on brown dwarf atmospheres is one area of improvement that has yet to be fully explored in SSEMs. The presence of additional metals could affect the amount of cloud formation and which condensates are formed, both of which would affect the opacity of the atmosphere and therefore the observed luminosity of the brown dwarf \citep{mar15}. Future SSEMs such as the Sonora models \citep{mar17} plan to cover a wider range of metallicities.

To improve the comparisons of HD~4747~B and HD~19467~B to SSEMs, more study should be done to constrain the masses and the ages of the brown dwarfs. Mass estimates will be improved with more radial velocity and direct imaging data combined with the latest parallaxes from Gaia DR2 \citep{bra18}. Current age estimates are highly disparate and method-dependent. Although neither HD~4747~A nor HD~19467~A are on the TASC target list, both stars should be targeted with the TESS two-minute cadence and could be studied with asteroseismology to help resolve the age discrepancy.


\acknowledgements
We would like to extend our sincere gratitude to Chris Farrington, Olli Majoinen, and Norm Vargas for CHARA observing support. We also thank the referee for their thorough and helpful review, Jamie Tayar for providing discussion and references for weakened magnetic breaking, and Patrick Fasano for help debugging and streamlining Python code.
This research made use of the SIMBAD and VIZIER Astronomical Databases, operated at CDS, Strasbourg, France (http://cdsweb.u-strasbg.fr/), of NASA's Astrophysics Data System, and of the Jean-Marie Mariotti Center's JSDC catalogue (http://www.jmmc.fr/catalogue\_jsdc.htm), co-developed by FIZEAU and LAOG/IPAG.
This work is based upon observations obtained with the Georgia State University Center for High Angular Resolution Astronomy Array at Mount Wilson Observatory.  The CHARA Array is supported by the National Science Foundation under Grant No. AST-1211929, AST-1636624, and AST-1715788.  Institutional support has been provided from the GSU College of Arts and Sciences and the GSU Office of the Vice President for Research and Economic Development.
C. M. W. acknowledges support from the Arthur J. Schmitt Leadership Fellowship at the University of Notre Dame.
T. B. acknowledges support provided through NASA grants ADAP12-0172, 14-XRP14\_2-0147, and 15-K2GO3\_2-0063.
K. v. B. acknowledges support provided through NASA/JPL grant RSA 1523106.
J. M. B. gratefully acknowledges support from NSF grant 1616086.
J. R. C. acknowledges support from the NASA Early Career program and the NSF CAREER program.
T. R. W. acknowledges the support of the Villum Foundation research grant 10118.

\facility{The CHARA Array, Keck HIRES}

\software{{\tt corner} \citep{for16}, {\tt emcee} \citep{for13}, {\tt isochrones} \citep{mor15}, MultiNest \citep{fer08, fer09, fer13}, SciPy \citep{jon01}}


\end{document}